\documentclass[%
 superscriptaddress,
 reprint,
 amsmath,
 amssymb,
 prl,
 showpacs,
 twocolumn,
 floatfix,
 nofootinbib,
]{revtex4-2}

\usepackage{graphicx}
\usepackage{dcolumn}
\usepackage{bm}
\usepackage{hyperref}
\usepackage[mathlines]{lineno}
\usepackage{color}

\begin{document}

\title{Light Dark Matter Search with a High-Resolution Athermal Phonon Detector Operated Above Ground}

\author{I.~Alkhatib} \affiliation{Department of Physics, University of Toronto, Toronto, ON M5S 1A7, Canada}
\author{D.W.P.~Amaral} \affiliation{Department of Physics, Durham University, Durham DH1 3LE, UK}
\author{T.~Aralis} \affiliation{Division of Physics, Mathematics, \& Astronomy, California Institute of Technology, Pasadena, CA 91125, USA}
\author{T.~Aramaki} \affiliation{SLAC National Accelerator Laboratory/Kavli Institute for Particle Astrophysics and Cosmology, Menlo Park, CA 94025, USA}
\author{I.J.~Arnquist} \affiliation{Pacific Northwest National Laboratory, Richland, WA 99352, USA}
\author{I.~Ataee~Langroudy} \affiliation{Department of Physics and Astronomy, and the Mitchell Institute for Fundamental Physics and Astronomy, Texas A\&M University, College Station, TX 77843, USA}
\author{E.~Azadbakht} \affiliation{Department of Physics and Astronomy, and the Mitchell Institute for Fundamental Physics and Astronomy, Texas A\&M University, College Station, TX 77843, USA}
\author{S.~Banik} \affiliation{School of Physical Sciences, National Institute of Science Education and Research, HBNI, Jatni - 752050, India}
\author{D.~Barker} \affiliation{School of Physics \& Astronomy, University of Minnesota, Minneapolis, MN 55455, USA}
\author{C.~Bathurst} \affiliation{Department of Physics, University of Florida, Gainesville, FL 32611, USA}
% ...
\collaboration{SuperCDMS Collaboration}
\author{J.~Camilleri} \altaffiliation[Now at ]{Department of Physics, Virginia Tech, Blacksburg, VA 24061, USA} \affiliation{Department of Physics, University of California, Berkeley, CA 94720, USA}
\author{Yu.G.~Kolomensky} \affiliation{Department of Physics, University of California, Berkeley, CA 94720, USA}\affiliation{Lawrence Berkeley National Laboratory, Berkeley, CA 94720, USA}
\author{S.~Zuber} \affiliation{Department of Physics, University of California, Berkeley, CA 94720, USA}

\date{\today}

\begin{abstract}
We present limits on spin-independent dark matter-nucleon interactions using a $10.6\,\mathrm{g}$ Si athermal phonon detector with a baseline energy resolution of ${\sigma_E = 3.86 \pm 0.04 \, (\mathrm{stat.})^{+0.19}_{-0.00} \, (\mathrm{syst.}) \, \mathrm{eV}}$. This exclusion analysis sets the most stringent dark matter-nucleon scattering cross-section limits achieved by a cryogenic detector for dark matter particle masses from $93$ to $140\,\mathrm{MeV}/c^2$, with a raw exposure of $9.9\,\mathrm{g\,d}$ acquired at an above-ground facility. This work illustrates the scientific potential of detectors with athermal phonon sensors with eV-scale energy resolution for future dark matter searches.
\end{abstract}

\maketitle


\textit{Introduction.---}Numerous observations have shown that the majority of the Universe is composed of nonluminous matter~\cite{pdg, planck,dm_evidence}. The weakly-interacting massive particle (WIMP)~\cite{wimp} has long been a favored candidate for this dark matter (DM). However, direct detection experiments have ruled out a significant portion of the most compelling WIMP parameter space~\cite{xenon1t_wimp,PhysRevLett.118.021303,PhysRevLett.119.181302}, which has motivated both theoretical and experimental exploration of alternative DM models~\cite{battaglieri2017cosmic}. In particular, light dark matter (LDM) with a mass in the $\mathrm{keV}/c^2$ to $\mathrm{GeV}/c^2$ range and coupling to Standard Model particles via a new force mediator provides a well-motivated alternative to the WIMP hypothesis~\cite{subgev,dark2013,dark2016}. While recent LDM searches have focused on DM-electron interactions~\cite{hvev,sensei,damic,xenon10,xenon1t_erdm}, detectors with eV-scale energy thresholds can also be used to study LDM via DM-nucleon interactions.

We present results from a DM search with a new Cryogenic PhotoDetector (CPD) featuring an athermal phonon sensor with a baseline energy resolution of ${\sigma_E = 3.86 \pm 0.04 \, (\mathrm{stat.})^{+0.19}_{-0.00} \, (\mathrm{syst.}) \, \mathrm{eV}}$. Although this device was designed for active particle identification in rare event searches~\cite{cpdcollaboration2020performance}, such as for neutrinoless double-beta decay~\cite{alphas,group2019cupid} and DM, the excellent energy resolution motivated its use as a DM detector itself. As a combined effort of the SuperCDMS and CPD collaborations, a DM search was carried out with $9.9\,\mathrm{g\,d}$ of raw exposure from Sept. $9^\mathrm{th}$ to $10^\mathrm{th}$ 2018. The data were acquired at the SLAC National Accelerator Laboratory in a surface facility of $\sim\!100\,\mathrm{m}$ in elevation. We discuss data acquisition techniques, device performance, and the results of an exclusion analysis for spin-independent DM-nucleon interactions.


\textit{Experimental setup.---}The CPD substrate is a $1\,\mathrm{mm}$ thick Si wafer with a radius of $3.81\,\mathrm{cm}$ and a mass of $10.6\,\mathrm{g}$. It is instrumented on one side with ${\sim\!1000}$ Quasiparticle-trap-assisted Electrothermal feedback Transition-edge sensors (QETs)~\cite{irwin,qet} distributed over the surface and connected in parallel to a single readout channel. The opposite side of the wafer is unpolished and not instrumented. The distributed channel results in minimal position dependence and fast collection of athermal phonons, which reduces inefficiency due to effects such as athermal phonon down-conversion~\cite{downconversion,knaak}. The eV-scale baseline energy resolution was achieved in part because of the relatively low QET critical temperature of $41.5\,\mathrm{mK}$ with a nominal bath temperature of $8\,\mathrm{mK}$.

A collimated $^{55}\mathrm{Fe}$ source was placed facing the noninstrumented side. The electron capture decay provides Mn $\mathrm{K}_\alpha$ and $\mathrm{K}_\beta$ X-ray lines at $5.9$ and $6.5\,\mathrm{keV}$, respectively, for \textit{in situ} calibration~\cite{fe55}. A $38\,\mu\mathrm{m}$ layer of Al was placed in front of the collimator to attenuate the rate of these photons and provide an additional calibration line at $1.5\,\mathrm{keV}$ from Al fluorescence~\cite{alfluor}.

For the sensor readout, a direct-current superconducting quantum interference device (SQUID) array--based amplifier was used, similar in design to the one described in Ref.~\cite{revc}. Because of project time constraints and large cosmogenic backgrounds, the DM search was limited to $22\,\mathrm{hrs}$. Data were acquired over this period using a field-programmable gate array (FPGA) triggering algorithm based on the optimal filter (OF) formalism~\cite{OF,golwala}. Throughout the exposure, randomly triggered samples of the baseline noise were acquired (``in-run random triggers"), which allowed us to observe any changes in the noise over the course of the search and to calculate and monitor the baseline energy resolution.

The trigger threshold was set at $4.2\,\sigma$ above the normally-distributed baseline noise level, corresponding to $16.3\,\mathrm{eV}$ after calibration. The phonon-pulse template used for the FPGA triggering algorithm was a double-exponential pulse with a rise time of $\tau_r = 20 \, \mu\mathrm{s}$ and a fall time of $\tau_f = 58\, \mu\mathrm{s}$. The rise time was taken from the expected collection time of athermal phonons, and the fall time was taken from the thermal response time of the QET estimated from a measurement of the complex admittance~\cite{irwin}. Each of these time constants was confirmed by a nonlinear least squares fit to nonsaturated pulses. Although Ref.~\cite{cpdcollaboration2020performance} discusses the existence of extra fall times, their effect on the OF amplitude measured for each event is negligible. Before starting the DM search, a separate, small subset of random triggers was collected. After removing data contaminated by effects such as elevated baselines and phonon pulses, the noise spectrum used by the FPGA algorithm was generated from these random triggers.

For overlapping triggered pulse traces, the triggering algorithm was set to save a trace centered on the pulse with the largest OF amplitude. We note that the FPGA triggering algorithm acted on a trace that was downsampled by a factor of 16, from the digitization rate of $625\,\mathrm{kHz}$ to $39\,\mathrm{kHz}$. Additionally, the FPGA triggering algorithm considered only $26.2\,\mathrm{ms}$ of the total $52\,\mathrm{ms}$ long time trace saved for each triggered pulse trace (``event"). Because of these factors, the energy resolution of the FPGA triggering algorithm is not as good as can be achieved by reconstructing event energies using an offline OF, as described in the following sections.

\begin{figure}
\includegraphics[width=\linewidth]{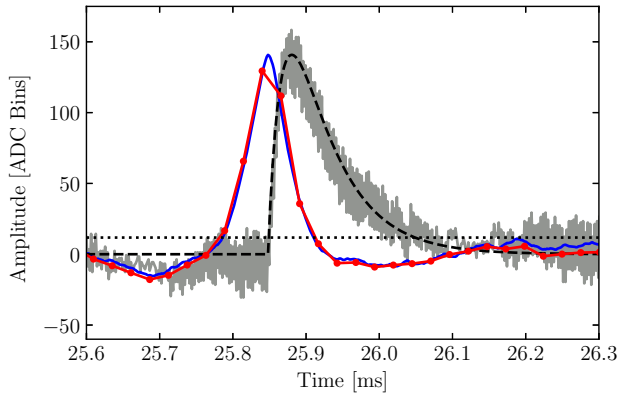}
\caption{\label{fig:event}A zoomed-in portion of an example event within the analysis ROI. The raw pulse (gray) is compared to the offline optimal filter result (blue), the pulse template scaled by the fit result (black dashed), the FPGA filter result (red with dots), and the FPGA trigger threshold (black dotted). The offline and FPGA optimal filters are highly correlated, but not exactly the same, with corresponding energy estimates for this event of $187\,\mathrm{eV}$ and $179\,\mathrm{eV}$, respectively. The offset between the optimal filters and the raw pulse is an artifact of the filters, as they were set up to determine the time of the beginning of the pulse, as opposed to the maximum of the pulse.}
\end{figure}

While the FPGA-based OF was used to trigger the experiment in real time, we ultimately used an offline algorithm to reconstruct event energies, where we again used the OF formalism. For this offline OF, we were able to use a single noise spectrum computed from the in-run random triggers to represent the entire data set because there was negligible time variation of the noise over the course of the full exposure. Pulse amplitudes and start times were reconstructed using the same phonon-pulse template as in the FPGA triggering algorithm. Thus, there are two different pulse amplitudes for each event---one from the FPGA triggering algorithm and one from the offline OF. In Fig.~\ref{fig:event}, we compare the different energy estimators for a representative event.

\begin{figure}
\includegraphics[width=\linewidth]{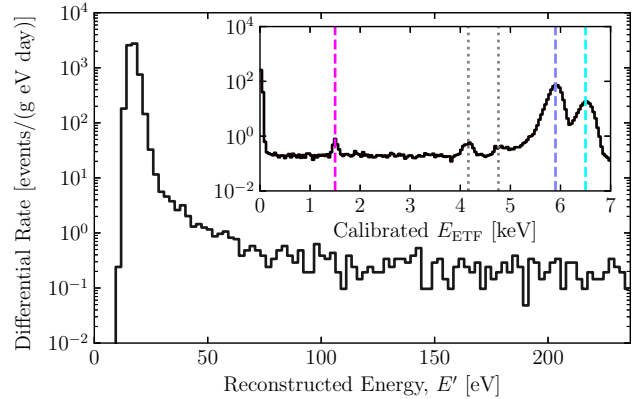}
\caption{\label{fig:spectrum}Measured energy spectrum in the DM-search ROI for the full exposure after application of the quality cuts. The data have been normalized to events per gram per day per eV and have been corrected for the event-selection efficiency, but not the trigger efficiency. The inset shows the calibrated $E_\mathrm{ETF}$ spectrum up to $7\,\mathrm{keV}$, noting the locations of the different spectral peaks. The known values of the dashed lines are $1.5$, $5.9$, and $6.5\,\mathrm{keV}$ for the Al fluorescence (pink), $^{55}\mathrm{Fe}$ $\mathrm{K}_\alpha$ (blue), and $^{55}\mathrm{Fe}$ $\mathrm{K}_\beta$ (cyan) lines, respectively. The two dotted gray lines between 4 and $5\,\mathrm{keV}$ in calibrated $E_\mathrm{ETF}$ are the Si escape peaks~\cite{Reed_1972}.}
\end{figure}

This detector was optimized for maximum energy sensitivity at low energies and does not have a large enough dynamic range to observe the calibration lines without nonlinear effects from saturation of the QETs. The nonlinearity is minimal within our region-of-interest (ROI), which is below $240 \, \mathrm{eV}$. Above the ROI, the fall time of the pulses increases monotonically with energy, which can be explained by effects of local saturation. Localized events can saturate nearby QETs to above the superconducting transition, while QETs far from the event stay within the superconducting transition. Because this is a single-channel device, the saturated and unsaturated QETs are read out in parallel and thus effectively combine into a single phonon pulse with an increased fall time. In order to correct out the saturation effects within the ROI, we follow the calibration method as outlined in Ref.~\cite{cpdcollaboration2020performance}. That is, the energy removed by electrothermal feedback ($E_\mathrm{ETF}$)~\cite{irwin} is saturation-corrected using an exponential model, and the OF-based energy estimators are converted to units of energy via a linear fit to the calibrated $E_\mathrm{ETF}$ within the ROI. With this method, there is an asymmetric systematic error in the baseline energy resolution, for which the upper bound corresponds to the value achieved when calibrating $E_\mathrm{ETF}$ linearly to the Al fluorescence line as opposed to the aforementioned exponential model (the lower bound). In Fig.~\ref{fig:spectrum}, we show the differential rate spectrum of the calibrated offline OF amplitude, with the inset showing the differential rate spectrum for the calibrated $E_\mathrm{ETF}$.

\textit{Data selection and efficiency.---}We make our final event selection with a minimal number of selection criteria (``cuts") to remove poorly reconstructed events without introducing energy dependence into the selection efficiency. This approach helps to reduce the complexity of the analysis and thus avoid introduction of systematic uncertainties. We apply two data-quality cuts: a prepulse baseline cut and a chi-square cut.

We define the event baseline as the average output in the prepulse section of each event, which is the first $25.6\,\mathrm{ms}$ of each trace. Large energy depositions have a long recovery time, which may manifest itself as a sloped baseline for subsequent events. Our trigger has reduced efficiency for any low-energy events occurring on such a baseline. We expected roughly 10\% of the events to sit on the tail of a high energy event in part because of the high muon flux at the surface of  $\sim\!1\,\mathrm{muon}/\mathrm{cm}^2/\mathrm{min}$~\cite{muonflux}. The baseline cut is performed by binning the data across the search in $400\,\mathrm{s}$ long bins and removing from each bin the 10\% of events that have the highest baseline.

The chi-square cut is a general cut on our goodness-of-fit metric, for which we use the low-frequency chi-square $\chi^2_\mathrm{LF}$ calculated from the offline OF fit~\cite{golwala}. This metric is similar to the $\chi^2$ from the offline OF fit, but we exclude frequencies over $f_\mathrm{cutoff}$ from the sum. This truncation allows us to remove sensitivity to superfluous degrees of freedom outside of our signal band from the chi-square, thereby reducing both the expected mean and the expected variance of the chi-square distribution. In this analysis, we used $f_\mathrm{cutoff} = 16\,\mathrm{kHz}$ because the rise and fall times of our expected pulse shape correspond to frequencies of $8.0\,\mathrm{kHz}$ and $2.7\,\mathrm{kHz}$ respectively. The pulse-shape variation within the DM-search ROI is minimal; this leads to a chi-square distribution that is largely independent of reconstructed event energy within this range. This in turn allows us to set an energy-independent cutoff value for $\chi^2_\mathrm{LF}$.

Our measured events cannot be used directly to measure the signal efficiency of the chi-square cut because they include some that are not representative of the expected DM signal, e.g. vibrationally-induced events, electronic glitches, pileup events, etc. Therefore, we created a pulse simulation by adding noise from the in-run random triggers to the pulse template, systematically scaling the latter over the range of energies corresponding to the DM-search ROI. We then process and analyze the simulated data in the same way as the DM-search data. In this case, the passage fraction of the chi-square cut, which has an energy-independent value of $98.53\pm0.01\%$, represents the cut's efficiency. 

We do not apply any other cuts to the DM-search data. The total signal efficiency is thus 88.7\%  and is independent of energy. A variation of the cut values within reasonable bounds was found to have no significant impact on the experimental sensitivity.


\textit{Signal model.---}In our DM signal model for spin-independent nuclear-recoil interactions~\cite{lewinsmith}, we use the standard astrophysical parameters for the dark matter velocity distribution~\cite{kerr,rave,schonrich}: a velocity of the Sun about the galactic center of $v_0 = 220\,\mathrm{km}/\mathrm{s}$, a mean orbital velocity of the Earth of $v_E = 232\,\mathrm{km}/\mathrm{s}$, a galactic escape velocity of $v_\mathrm{esc} = 544\,\mathrm{km}/\mathrm{s}$, and a local DM density of $\rho_0 = 0.3 \,\mathrm{GeV}/\mathrm{cm}^3$. To take into account the trigger efficiency, we convolve the differential rate with the joint probability density function relating our two energy estimators, including the effects of the applied cuts. The signal model, which includes the estimated trigger efficiency, is given by
\begin{eqnarray}
    {\frac{\partial R}{\partial E'}(E')}=&&\int_0^\infty \mathop{dE_T} \int_0^\infty  \mathop{dE_0} \Theta(E_T - \delta)\nonumber\\
    &&\times  \varepsilon(E', E_T, E_0) P(E',E_T|E_0)\frac{\partial R}{\partial E_0}(E_0) ,
    \label{eq:signal}
\end{eqnarray}
where $E_0$ is the true recoil energy, $E'$ is the recoil energy measured by the offline OF, $E_T$ is the recoil energy measured by the FPGA triggering algorithm, $\delta$ is the trigger threshold set on the FPGA triggering algorithm, $\varepsilon$ is the efficiency of the two quality cuts and two cuts that are applied to simulated data (as described in the following paragraphs), $\Theta$ represents the trigger threshold cut (a Heaviside function), and $P(E',E_T|E_0)$ is the probability to extract $E'$ and $E_T$ using the two energy reconstruction algorithms given the true recoil energy $E_0$. For the efficiency $\varepsilon$, we have generalized its form to be a function of energy, knowing that the baseline and chi-square cuts themselves are energy independent. The heat quenching factor (the ratio of heat signals produced by nuclear and electron recoils of the same energy that accounts for effects such as displacement damage) has been assumed to be unity for this work. Though measurements of the heat quenching factor have not been made for Si, similar work has been undertaken for Ge, where the heat quenching factor was shown to be very close to unity~\cite{BENOIT2007558,206Pb}.

The model in Eq.~(\ref{eq:signal}) was evaluated numerically, taking advantage of our pulse simulation. The pulse simulation includes a software simulation of the FPGA triggering algorithm, which had the same output as the hardware version when run on the DM-search data. With this simulation of the FPGA triggering algorithm, we can use the pulse simulation to determine $P(E',E_T|E_0)$ directly. Low pulse height events may have their OF energy estimate affected by a shift of the start time estimate, but the simulation automatically takes this effect into account.

We also added two cuts to the simulated data only: a confidence ellipse cut and a trigger time cut. The confidence ellipse cut removed any events with an energy estimator value outside of the 99.7\% confidence ellipse, which is defined by the covariance matrix of our two energy estimators for zero-energy events. This cut was implemented to exclude the possible scenario of calculating a finite sensitivity to zero-energy DM, which would be a nonphysical result. The trigger time cut removed events that were not within $29\,\mu \mathrm{s}$---half of a fall time of a pulse---of the true event time, as determined by the energy-scaled pulse template. This cut ensured that the triggering algorithm was able to detect the signal added, as opposed to a large noise fluctuation elsewhere in the trace. These two cuts required knowledge of the true energy of the pulse---they cannot be applied to the data, but can be applied to the signal model---and helped to ensure that our signal modeling was conservative. In adding each of these cuts, we reduced our signal efficiency estimate, which necessarily biased the results in the conservative direction.


\textit{Results.---}The objective of this DM search was to set conservative limits on the spin-independent interaction of dark matter particles with masses below $1.5\,\mathrm{GeV}/c^2$. For the lower edge of the limit contour, we use the optimum interval (OI) method~\cite{yellin_upper, yellin_extended} with unknown background. For the upper edge of the limit contour, we use a modified version of the publicly available \textsc{verne} code~\cite{verne}, which uses a Poisson method to calculate the effects of overburden~\cite{starkman1990, zaharijas2005, overburden} on the DM signal. This code has been similarly used in Refs.~\cite{PhysRevD.97.123013, PhysRevLett.123.241803, edelweiss}. For the overburden assumption, we include the $5\,\mathrm{cm}$ of Cu surrounding the detector, the shielding from the atmosphere, and the shielding from the Earth. Both limit-setting methods assume that the full measured event rate could be due to a DM signal and set the limits at the 90\% confidence level (C.L.).

\begin{figure}
\includegraphics[width=\linewidth]{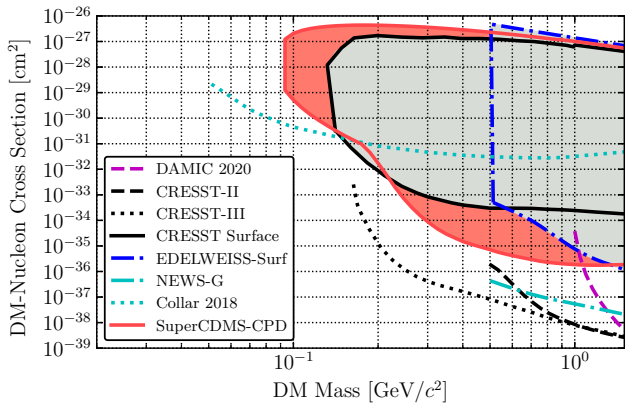}
\caption{\label{fig:limit}The 90\% C.L. limits on the spin-independent DM-nucleon cross section as a function of DM mass for this work (solid red line), compared to results from other experiments~\cite{edelweiss,cresst2019,cresst_aboveground,PhysRevLett.125.241803,cresst2,newsg,PhysRevD.98.023005}. For above-ground experiments with overburden calculations, the previously ruled out parameter space is shown as the gray shaded region, and the new parameter space ruled out from this search is shown as the red shaded region. For the Collar 2018 surface result, which uses a liquid scintillator cell operated at $1^{\circ} \, \mathrm{C}$, an overburden calculation would be useful for comparison to the upper edges of the various contours for the surface searches. We note that the systematic error in the baseline energy resolution changes the result within the error of the limit's line width, thus we only include the result from the $3.86\,\mathrm{eV}$ calibration.}
\end{figure}

The results of the dark matter search are shown in Fig.~\ref{fig:limit}, compared to other pertinent DM searches in the same parameter space~\cite{PhysRevLett.125.241803,cresst2,cresst2019,cresst_aboveground,edelweiss,newsg,PhysRevD.98.023005}. For DM masses between $93$ and $140\,\mathrm{MeV}/c^2$, these results provide the most stringent limits for nuclear-recoil DM signals using a cryogenic detector. For DM masses between $220\,\mathrm{MeV}/c^2$ and $1.35\,\mathrm{GeV}/c^2$, they are the most stringent limits achieved in an above-ground facility. For these low DM masses, the large cross sections approach the level at which the Born approximation used in the standard DM signal model begins to fail~\cite{PhysRevD.100.063013}. However, in the absence of a generally accepted alternative model and to be comparable to other experiments (all of which also use the Born approximation in this regime), we decided to keep it in our signal model as well.

To estimate the systematic error in the limit contour, we compared the results obtained by calculating the signal model using eight different sets of pulse simulations. The variation in the limit was found to be on the order of 10\% for DM masses below $200\, \mathrm{MeV}/c^2$ for the lower edge and below $100\, \mathrm{MeV}/c^2$ for the upper edge. Above these DM masses, the variation in each edge decreased to less than 1\%. The $\mathcal{O}(10\%)$ variation observed at the smallest DM masses is attributed to a greater uncertainty in the trigger efficiency for sub-threshold events, as opposed to events that are reconstructed with energies above threshold. In the limit shown in Fig.~\ref{fig:limit}, we have taken the median of the limits calculated for the eight simulations at each DM mass. The 10\% variation is not plotted, as it would not be visible in the figure.

\begin{figure}
\includegraphics[width=\linewidth]{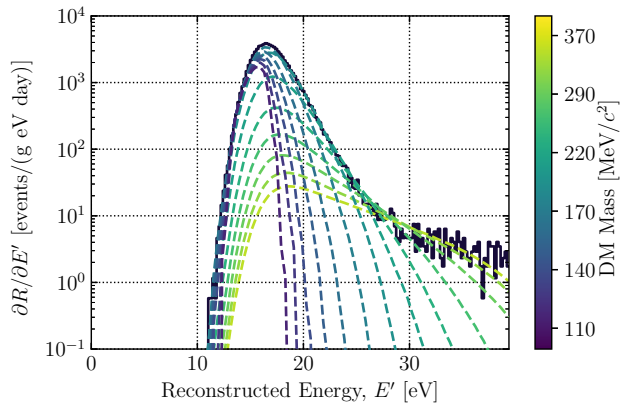}
\caption{\label{fig:drde}The event spectrum for the DM search data below $40\,\mathrm{eV}$ in reconstructed energy. The data have been normalized to events per gram per day per eV and have been corrected for the signal efficiency of the data-quality cuts, but not the confidence ellipse and trigger time cuts. The colored dashed lines represent the calculated event rates for selected DM cross sections and masses from the 90\% C.L. OI limit for a single pulse simulation, where the optimum intervals in recoil energy are below $40\,\mathrm{eV}$. Sensitivity to DM masses below $400\,\mathrm{MeV}/c^2$ corresponds to recoil energies below $40\,\mathrm{eV}$, with the lowest masses requiring energy sensitivity down to about $15\,\mathrm{eV}$.}
\end{figure}

In Fig.~\ref{fig:drde}, we show the data spectrum for reconstructed energies below $40\,\mathrm{eV}$ and DM signal curves for various DM masses for a single pulse simulation, where the cross sections from the OI limit are used. The approximate location of the optimum interval is apparent for each dark matter mass.

In this search, we see an excess of events for recoil energies below about $100\,\mathrm{eV}$, emerging above the roughly flat rate from Compton scattering of the gamma-ray background. This excess of events could be from an unknown external background or due to detector effects such as crystal cracking~\cite{ASTROM2006262}. As other experiments have observed excess events in searches for low-mass nuclear-recoiling DM~\cite{cresst2019, cresst_aboveground, edelweiss, PhysRevD.102.015017}, understanding this background is of pivotal importance. Future studies will be devoted to this, and we are actively investigating this excess by operating this detector in an environment with substantially reduced cosmogenic backgrounds.

\textit{Conclusion.---}Using a detector with $\sigma_E = 3.86 \pm 0.04 \, (\mathrm{stat.})^{+0.19}_{-0.00} \, (\mathrm{syst.}) \, \mathrm{eV}$ baseline energy resolution operated in an above-ground facility with an exposure of $9.9\,\mathrm{g\,d}$, we probe parameter space for spin-independent interactions of DM with nucleons for dark matter particles with masses above $93\,\mathrm{MeV}/c^2$. The range from $93$ to $140\,\mathrm{MeV}/c^2$ was previously not accessible to cryogenic detectors. These results also set the most stringent limits for above-ground nuclear-recoil signals from dark matter for masses between $220\,\mathrm{MeV}/c^2$ and $1.35\,\mathrm{GeV}/c^2$. This was achieved using a single readout channel composed of QETs distributed on a Si substrate, with a recoil energy threshold set at $16.3\,\mathrm{eV}$.

The results of this work were accomplished despite the high background rates in our surface facility because of the excellent baseline energy resolution of the detector. We plan to operate this detector in an underground laboratory, where we expect to have a significantly lower Compton scattering background rate. This will allow further study of the excess events observed in the ROI, hopefully providing insight into the origin of the event rate that is limiting the results reported here.

These results also demonstrate the potential of athermal phonon sensors with eV-scale baseline energy resolution for future dark matter searches via DM-nucleon interactions. Because this detector has a large surface area relative to its small volume, it is not optimal for a DM search. The baseline energy resolution of such devices scales with the number of QETs, which itself is proportional to the instrumented area (assuming the same QET design used by the CPD)~\cite{Hochberg_2016,KNAPEN2018386}. Thus, a decrease in the instrumented area, with an increase in volume, should lead to improvements in baseline energy resolution. Future work is planned to design detectors of volume $\sim\!1\,\mathrm{cm}^3$, for which it is reasonable to expect roughly an order of magnitude improvement in baseline energy resolution through these geometric considerations alone. With improved baseline energy resolution comes a lower energy threshold, allowing a search for spin-independent DM-nucleon interactions for even lower DM masses and a clear path to surpassing the existing noncryogenic detector constraints on sub-$100\, \mathrm{MeV}/c^2$ DM interacting with nucleons.

\textit{Acknowledgments.---}Funding and support were received from the National Science Foundation, the U.S. Department of Energy (DOE), Fermilab URA Visiting Scholar Grant No. 15-S-33, NSERC Canada, the Canada First Excellence Research  Fund, the Arthur B. McDonald Institute (Canada), Michael M. Garland, the Department of Atomic Energy Government of India (DAE), and the Deutsche Forschungsgemeinschaft (DFG, German Research Foundation)---Project No. 420484612 and under Germany's Excellence Strategy - EXC 2121 ``Quantum Universe" -- 390833306. Fermilab, PNNL, SLAC, and LBNL are operated under contracts DE-AC02-07CH11359, DE-AC05-76RL01830, DE-AC02-76SF00515, and DE-AC02-05CH11231, respectively, with the U.S. Department of Energy.

\providecommand{\noopsort}[1]{}\providecommand{\singleletter}[1]{#1}%


\begin{thebibliography}{56}%
\makeatletter
\providecommand \@ifxundefined [1]{%
 \@ifx{#1\undefined}
}%
\providecommand \@ifnum [1]{%
 \ifnum #1\expandafter \@firstoftwo
 \else \expandafter \@secondoftwo
 \fi
}%
\providecommand \@ifx [1]{%
 \ifx #1\expandafter \@firstoftwo
 \else \expandafter \@secondoftwo
 \fi
}%
\providecommand \natexlab [1]{#1}%
\providecommand \enquote  [1]{``#1''}%
\providecommand \bibnamefont  [1]{#1}%
\providecommand \bibfnamefont [1]{#1}%
\providecommand \citenamefont [1]{#1}%
\providecommand \href@noop [0]{\@secondoftwo}%
\providecommand \href [0]{\begingroup \@sanitize@url \@href}%
\providecommand \@href[1]{\@@startlink{#1}\@@href}%
\providecommand \@@href[1]{\endgroup#1\@@endlink}%
\providecommand \@sanitize@url [0]{\catcode `\\12\catcode `\$12\catcode
  `\&12\catcode `\#12\catcode `\^12\catcode `\_12\catcode `\%12\relax}%
\providecommand \@@startlink[1]{}%
\providecommand \@@endlink[0]{}%
\providecommand \url  [0]{\begingroup\@sanitize@url \@url }%
\providecommand \@url [1]{\endgroup\@href {#1}{\urlprefix }}%
\providecommand \urlprefix  [0]{URL }%
\providecommand \Eprint [0]{\href }%
\providecommand \doibase [0]{https://doi.org/}%
\providecommand \selectlanguage [0]{\@gobble}%
\providecommand \bibinfo  [0]{\@secondoftwo}%
\providecommand \bibfield  [0]{\@secondoftwo}%
\providecommand \translation [1]{[#1]}%
\providecommand \BibitemOpen [0]{}%
\providecommand \bibitemStop [0]{}%
\providecommand \bibitemNoStop [0]{.\EOS\space}%
\providecommand \EOS [0]{\spacefactor3000\relax}%
\providecommand \BibitemShut  [1]{\csname bibitem#1\endcsname}%
\let\auto@bib@innerbib\@empty
\bibitem [{\citenamefont {Tanabashi}\ \emph {et~al.}(2018)\citenamefont
  {Tanabashi} \emph {et~al.}}]{pdg}%
  \BibitemOpen
  \bibfield  {author} {\bibinfo {author} {\bibfnamefont {M.}~\bibnamefont
  {Tanabashi}} \emph {et~al.} (\bibinfo {collaboration} {Particle Data
  Group}),\ }\bibinfo {title} {Review of Particle Physics},\ \href
  {https://doi.org/10.1103/PhysRevD.98.030001} {\bibfield  {journal} {\bibinfo
  {journal} {Phys. Rev. D}\ }\textbf {\bibinfo {volume} {98}},\ \bibinfo
  {pages} {030001} (\bibinfo {year} {2018})}\BibitemShut {NoStop}%
\bibitem [{\citenamefont {Ade}\ \emph {et~al.}(2016)\citenamefont {Ade} \emph
  {et~al.}}]{planck}%
  \BibitemOpen
  \bibfield  {author} {\bibinfo {author} {\bibfnamefont {P.~A.~R.}\
  \bibnamefont {Ade}} \emph {et~al.} (\bibinfo {collaboration} {Planck
  Collaboration}),\ }\bibinfo {title} {Planck 2015 results - XIII. Cosmological
  parameters},\ \href {https://doi.org/10.1051/0004-6361/201525830} {\bibfield
  {journal} {\bibinfo  {journal} {Astron. Astrophys.}\ }\textbf {\bibinfo
  {volume} {594}},\ \bibinfo {pages} {A13} (\bibinfo {year}
  {2016})}\BibitemShut {NoStop}%
\bibitem [{\citenamefont {Clowe}\ \emph {et~al.}(2006)\citenamefont {Clowe},
  \citenamefont {Brada{\v{c}}}, \citenamefont {Gonzalez}, \citenamefont
  {Markevitch}, \citenamefont {Randall}, \citenamefont {Jones},\ and\
  \citenamefont {Zaritsky}}]{dm_evidence}%
  \BibitemOpen
  \bibfield  {author} {\bibinfo {author} {\bibfnamefont {D.}~\bibnamefont
  {Clowe}}, \bibinfo {author} {\bibfnamefont {M.}~\bibnamefont {Brada{\v{c}}}},
  \bibinfo {author} {\bibfnamefont {A.~H.}\ \bibnamefont {Gonzalez}}, \bibinfo
  {author} {\bibfnamefont {M.}~\bibnamefont {Markevitch}}, \bibinfo {author}
  {\bibfnamefont {S.~W.}\ \bibnamefont {Randall}}, \bibinfo {author}
  {\bibfnamefont {C.}~\bibnamefont {Jones}},\ and\ \bibinfo {author}
  {\bibfnamefont {D.}~\bibnamefont {Zaritsky}},\ }\bibinfo {title} {A Direct
  Empirical Proof of the Existence of Dark Matter},\ \href
  {https://doi.org/10.1086/508162} {\bibfield  {journal} {\bibinfo  {journal}
  {Astrophys. J.}\ }\textbf {\bibinfo {volume} {648}},\ \bibinfo {pages} {L109}
  (\bibinfo {year} {2006})}\BibitemShut {NoStop}%
\bibitem [{\citenamefont {Steigman}\ and\ \citenamefont {Turner}(1985)}]{wimp}%
  \BibitemOpen
  \bibfield  {author} {\bibinfo {author} {\bibfnamefont {G.}~\bibnamefont
  {Steigman}}\ and\ \bibinfo {author} {\bibfnamefont {M.~S.}\ \bibnamefont
  {Turner}},\ }\bibinfo {title} {Cosmological constraints on the properties of
  weakly interacting massive particles},\ \href
  {https://doi.org/10.1016/0550-3213(85)90537-1} {\bibfield  {journal}
  {\bibinfo  {journal} {Nucl. Phys. B}\ }\textbf {\bibinfo {volume} {253}},\
  \bibinfo {pages} {375 } (\bibinfo {year} {1985})}\BibitemShut {NoStop}%
\bibitem [{\citenamefont {Aprile}\ \emph {et~al.}(2018)\citenamefont {Aprile}
  \emph {et~al.}}]{xenon1t_wimp}%
  \BibitemOpen
  \bibfield  {author} {\bibinfo {author} {\bibfnamefont {E.}~\bibnamefont
  {Aprile}} \emph {et~al.} (\bibinfo {collaboration} {XENON Collaboration}),\
  }\bibinfo {title} {Dark Matter Search Results from a One Ton-Year Exposure of
  XENON1T},\ \href {https://doi.org/10.1103/PhysRevLett.121.111302} {\bibfield
  {journal} {\bibinfo  {journal} {Phys. Rev. Lett.}\ }\textbf {\bibinfo
  {volume} {121}},\ \bibinfo {pages} {111302} (\bibinfo {year}
  {2018})}\BibitemShut {NoStop}%
\bibitem [{\citenamefont {Akerib}\ \emph {et~al.}(2017)\citenamefont {Akerib}
  \emph {et~al.}}]{PhysRevLett.118.021303}%
  \BibitemOpen
  \bibfield  {author} {\bibinfo {author} {\bibfnamefont {D.~S.}\ \bibnamefont
  {Akerib}} \emph {et~al.} (\bibinfo {collaboration} {LUX Collaboration}),\
  }\bibinfo {title} {Results from a Search for Dark Matter in the Complete LUX
  Exposure},\ \href {https://doi.org/10.1103/PhysRevLett.118.021303} {\bibfield
   {journal} {\bibinfo  {journal} {Phys. Rev. Lett.}\ }\textbf {\bibinfo
  {volume} {118}},\ \bibinfo {pages} {021303} (\bibinfo {year}
  {2017})}\BibitemShut {NoStop}%
\bibitem [{\citenamefont {Cui}\ \emph {et~al.}(2017)\citenamefont {Cui} \emph
  {et~al.}}]{PhysRevLett.119.181302}%
  \BibitemOpen
  \bibfield  {author} {\bibinfo {author} {\bibfnamefont {X.}~\bibnamefont
  {Cui}} \emph {et~al.} (\bibinfo {collaboration} {PandaX-II Collaboration}),\
  }\bibinfo {title} {Dark Matter Results from 54-Ton-Day Exposure of PandaX-II
  Experiment},\ \href {https://doi.org/10.1103/PhysRevLett.119.181302}
  {\bibfield  {journal} {\bibinfo  {journal} {Phys. Rev. Lett.}\ }\textbf
  {\bibinfo {volume} {119}},\ \bibinfo {pages} {181302} (\bibinfo {year}
  {2017})}\BibitemShut {NoStop}%
\bibitem [{\citenamefont {Battaglieri}\ \emph {et~al.}()\citenamefont
  {Battaglieri} \emph {et~al.}}]{battaglieri2017cosmic}%
  \BibitemOpen
  \bibfield  {author} {\bibinfo {author} {\bibfnamefont {M.}~\bibnamefont
  {Battaglieri}} \emph {et~al.},\ }\href@noop {} {\bibinfo {title} {US Cosmic
  Visions: New Ideas in Dark Matter 2017: Community Report}},\ \Eprint
  {https://arxiv.org/abs/1707.04591} {arXiv:1707.04591} \BibitemShut {NoStop}%
\bibitem [{\citenamefont {Essig}\ \emph
  {et~al.}(2012{\natexlab{a}})\citenamefont {Essig}, \citenamefont {Mardon},\
  and\ \citenamefont {Volansky}}]{subgev}%
  \BibitemOpen
  \bibfield  {author} {\bibinfo {author} {\bibfnamefont {R.}~\bibnamefont
  {Essig}}, \bibinfo {author} {\bibfnamefont {J.}~\bibnamefont {Mardon}},\ and\
  \bibinfo {author} {\bibfnamefont {T.}~\bibnamefont {Volansky}},\ }\bibinfo
  {title} {Direct detection of sub-GeV dark matter},\ \href
  {https://doi.org/10.1103/PhysRevD.85.076007} {\bibfield  {journal} {\bibinfo
  {journal} {Phys. Rev. D}\ }\textbf {\bibinfo {volume} {85}},\ \bibinfo
  {pages} {076007} (\bibinfo {year} {2012}{\natexlab{a}})}\BibitemShut
  {NoStop}%
\bibitem [{\citenamefont {Essig}\ \emph {et~al.}()\citenamefont {Essig} \emph
  {et~al.}}]{dark2013}%
  \BibitemOpen
  \bibfield  {author} {\bibinfo {author} {\bibfnamefont {R.}~\bibnamefont
  {Essig}} \emph {et~al.},\ }\href@noop {} {\bibinfo {title} {Dark Sectors and
  New, Light, Weakly-Coupled Particles}},\ \Eprint
  {https://arxiv.org/abs/1311.0029} {arXiv:1311.0029} \BibitemShut {NoStop}%
\bibitem [{\citenamefont {Alexander}\ \emph {et~al.}()\citenamefont {Alexander}
  \emph {et~al.}}]{dark2016}%
  \BibitemOpen
  \bibfield  {author} {\bibinfo {author} {\bibfnamefont {J.}~\bibnamefont
  {Alexander}} \emph {et~al.},\ }\href@noop {} {\bibinfo {title} {Dark Sectors
  2016 Workshop: Community Report}},\ \Eprint
  {https://arxiv.org/abs/1608.08632} {arXiv:1608.08632} \BibitemShut {NoStop}%
\bibitem [{\citenamefont {Agnese}\ \emph
  {et~al.}(2018{\natexlab{a}})\citenamefont {Agnese} \emph {et~al.}}]{hvev}%
  \BibitemOpen
  \bibfield  {author} {\bibinfo {author} {\bibfnamefont {R.}~\bibnamefont
  {Agnese}} \emph {et~al.} (\bibinfo {collaboration} {SuperCDMS
  Collaboration}),\ }\bibinfo {title} {First Dark Matter Constraints from a
  SuperCDMS Single-Charge Sensitive Detector},\ \href
  {https://doi.org/10.1103/PhysRevLett.121.051301} {\bibfield  {journal}
  {\bibinfo  {journal} {Phys. Rev. Lett.}\ }\textbf {\bibinfo {volume} {121}},\
  \bibinfo {pages} {051301} (\bibinfo {year} {2018}{\natexlab{a}})}\BibitemShut
  {NoStop}%
\bibitem [{\citenamefont {Abramoff}\ \emph {et~al.}(2019)\citenamefont
  {Abramoff} \emph {et~al.}}]{sensei}%
  \BibitemOpen
  \bibfield  {author} {\bibinfo {author} {\bibfnamefont {O.}~\bibnamefont
  {Abramoff}} \emph {et~al.} (\bibinfo {collaboration} {SENSEI
  Collaboration}),\ }\bibinfo {title} {SENSEI: Direct-Detection Constraints on
  Sub-GeV Dark Matter from a Shallow Underground Run Using a Prototype Skipper
  CCD},\ \href {https://doi.org/10.1103/PhysRevLett.122.161801} {\bibfield
  {journal} {\bibinfo  {journal} {Phys. Rev. Lett.}\ }\textbf {\bibinfo
  {volume} {122}},\ \bibinfo {pages} {161801} (\bibinfo {year}
  {2019})}\BibitemShut {NoStop}%
\bibitem [{\citenamefont {Aguilar-Arevalo}\ \emph {et~al.}(2019)\citenamefont
  {Aguilar-Arevalo} \emph {et~al.}}]{damic}%
  \BibitemOpen
  \bibfield  {author} {\bibinfo {author} {\bibfnamefont {A.}~\bibnamefont
  {Aguilar-Arevalo}} \emph {et~al.} (\bibinfo {collaboration} {DAMIC
  Collaboration}),\ }\bibinfo {title} {Constraints on Light Dark Matter
  Particles Interacting with Electrons from DAMIC at SNOLAB},\ \href
  {https://doi.org/10.1103/PhysRevLett.123.181802} {\bibfield  {journal}
  {\bibinfo  {journal} {Phys. Rev. Lett.}\ }\textbf {\bibinfo {volume} {123}},\
  \bibinfo {pages} {181802} (\bibinfo {year} {2019})}\BibitemShut {NoStop}%
\bibitem [{\citenamefont {Essig}\ \emph
  {et~al.}(2012{\natexlab{b}})\citenamefont {Essig}, \citenamefont
  {Manalaysay}, \citenamefont {Mardon}, \citenamefont {Sorensen},\ and\
  \citenamefont {Volansky}}]{xenon10}%
  \BibitemOpen
  \bibfield  {author} {\bibinfo {author} {\bibfnamefont {R.}~\bibnamefont
  {Essig}}, \bibinfo {author} {\bibfnamefont {A.}~\bibnamefont {Manalaysay}},
  \bibinfo {author} {\bibfnamefont {J.}~\bibnamefont {Mardon}}, \bibinfo
  {author} {\bibfnamefont {P.}~\bibnamefont {Sorensen}},\ and\ \bibinfo
  {author} {\bibfnamefont {T.}~\bibnamefont {Volansky}},\ }\bibinfo {title}
  {First Direct Detection Limits on Sub-GeV Dark Matter from XENON10},\ \href
  {https://doi.org/10.1103/PhysRevLett.109.021301} {\bibfield  {journal}
  {\bibinfo  {journal} {Phys. Rev. Lett.}\ }\textbf {\bibinfo {volume} {109}},\
  \bibinfo {pages} {021301} (\bibinfo {year} {2012}{\natexlab{b}})}\BibitemShut
  {NoStop}%
\bibitem [{\citenamefont {Aprile}\ \emph
  {et~al.}(2019{\natexlab{a}})\citenamefont {Aprile} \emph
  {et~al.}}]{xenon1t_erdm}%
  \BibitemOpen
  \bibfield  {author} {\bibinfo {author} {\bibfnamefont {E.}~\bibnamefont
  {Aprile}} \emph {et~al.} (\bibinfo {collaboration} {XENON Collaboration}),\
  }\bibinfo {title} {Light Dark Matter Search with Ionization Signals in
  XENON1T},\ \href {https://doi.org/10.1103/PhysRevLett.123.251801} {\bibfield
  {journal} {\bibinfo  {journal} {Phys. Rev. Lett.}\ }\textbf {\bibinfo
  {volume} {123}},\ \bibinfo {pages} {251801} (\bibinfo {year}
  {2019}{\natexlab{a}})}\BibitemShut {NoStop}%
\bibitem [{\citenamefont {Fink}\ \emph {et~al.}(2021)\citenamefont {Fink} \emph
  {et~al.}}]{cpdcollaboration2020performance}%
  \BibitemOpen
  \bibfield  {author} {\bibinfo {author} {\bibfnamefont {C.~W.}\ \bibnamefont
  {Fink}} \emph {et~al.} (\bibinfo {collaboration} {CPD Collaboration}),\
  }\bibinfo {title} {Performance of a large area photon detector for rare event
  search applications},\ \href {https://doi.org/10.1063/5.0032372} {\bibfield
  {journal} {\bibinfo  {journal} {Appl. Phys. Lett.}\ }\textbf {\bibinfo
  {volume} {118}},\ \bibinfo {pages} {022601} (\bibinfo {year}
  {2021})}\BibitemShut {NoStop}%
\bibitem [{\citenamefont {Tabarelli~de Fatis}(2010)}]{alphas}%
  \BibitemOpen
  \bibfield  {author} {\bibinfo {author} {\bibfnamefont {T.}~\bibnamefont
  {Tabarelli~de Fatis}},\ }\bibinfo {title} {Cerenkov emission as a positive
  tag of double beta decays in bolometric experiments},\ \href
  {https://doi.org/10.1140/epjc/s10052-009-1207-8} {\bibfield  {journal}
  {\bibinfo  {journal} {Eur. Phys. J. C.}\ }\textbf {\bibinfo {volume} {65}},\
  \bibinfo {pages} {359} (\bibinfo {year} {2010})}\BibitemShut {NoStop}%
\bibitem [{\citenamefont {Armstrong}\ \emph {et~al.}()\citenamefont {Armstrong}
  \emph {et~al.}}]{group2019cupid}%
  \BibitemOpen
  \bibfield  {author} {\bibinfo {author} {\bibfnamefont {W.~R.}\ \bibnamefont
  {Armstrong}} \emph {et~al.} (\bibinfo {collaboration} {CUPID
  Collaboration}),\ }\href@noop {} {\bibinfo {title} {CUPID pre-CDR}},\ \Eprint
  {https://arxiv.org/abs/1907.09376} {arXiv:1907.09376} \BibitemShut {NoStop}%
\bibitem [{\citenamefont {Irwin}\ and\ \citenamefont {Hilton}(2005)}]{irwin}%
  \BibitemOpen
  \bibfield  {author} {\bibinfo {author} {\bibfnamefont {K.~D.}\ \bibnamefont
  {Irwin}}\ and\ \bibinfo {author} {\bibfnamefont {G.~C.}\ \bibnamefont
  {Hilton}},\ }\bibinfo {title} {Transition-Edge Sensors},\ in\ \href
  {https://doi.org/10.1007/10933596_3} {\emph {\bibinfo {booktitle} {Cryogenic
  Particle Detection}}},\ \bibinfo {editor} {edited by\ \bibinfo {editor}
  {\bibfnamefont {C.}~\bibnamefont {Enss}}}\ (\bibinfo  {publisher} {Springer
  Berlin Heidelberg},\ \bibinfo {address} {Berlin, Heidelberg},\ \bibinfo
  {year} {2005})\ pp.\ \bibinfo {pages} {63--150}\BibitemShut {NoStop}%
\bibitem [{\citenamefont {Irwin}\ \emph {et~al.}(1995)\citenamefont {Irwin},
  \citenamefont {Nam}, \citenamefont {Cabrera}, \citenamefont {Chugg},\ and\
  \citenamefont {Young}}]{qet}%
  \BibitemOpen
  \bibfield  {author} {\bibinfo {author} {\bibfnamefont {K.~D.}\ \bibnamefont
  {Irwin}}, \bibinfo {author} {\bibfnamefont {S.~W.}\ \bibnamefont {Nam}},
  \bibinfo {author} {\bibfnamefont {B.}~\bibnamefont {Cabrera}}, \bibinfo
  {author} {\bibfnamefont {B.}~\bibnamefont {Chugg}},\ and\ \bibinfo {author}
  {\bibfnamefont {B.~A.}\ \bibnamefont {Young}},\ }\bibinfo {title} {A
  quasiparticle‐trap‐assisted transition‐edge sensor for
  phonon‐mediated particle detection},\ \href
  {https://doi.org/10.1063/1.1146105} {\bibfield  {journal} {\bibinfo
  {journal} {Rev. Sci. Instrum.}\ }\textbf {\bibinfo {volume} {66}},\ \bibinfo
  {pages} {5322} (\bibinfo {year} {1995})}\BibitemShut {NoStop}%
\bibitem [{\citenamefont {Kozorezov}\ \emph {et~al.}(2000)\citenamefont
  {Kozorezov}, \citenamefont {Volkov}, \citenamefont {Wigmore}, \citenamefont
  {Peacock}, \citenamefont {Poelaert},\ and\ \citenamefont {den
  Hartog}}]{downconversion}%
  \BibitemOpen
  \bibfield  {author} {\bibinfo {author} {\bibfnamefont {A.~G.}\ \bibnamefont
  {Kozorezov}}, \bibinfo {author} {\bibfnamefont {A.~F.}\ \bibnamefont
  {Volkov}}, \bibinfo {author} {\bibfnamefont {J.~K.}\ \bibnamefont {Wigmore}},
  \bibinfo {author} {\bibfnamefont {A.}~\bibnamefont {Peacock}}, \bibinfo
  {author} {\bibfnamefont {A.}~\bibnamefont {Poelaert}},\ and\ \bibinfo
  {author} {\bibfnamefont {R.}~\bibnamefont {den Hartog}},\ }\bibinfo {title}
  {Quasiparticle-phonon downconversion in nonequilibrium superconductors},\
  \href {https://doi.org/10.1103/PhysRevB.61.11807} {\bibfield  {journal}
  {\bibinfo  {journal} {Phys. Rev. B}\ }\textbf {\bibinfo {volume} {61}},\
  \bibinfo {pages} {11807} (\bibinfo {year} {2000})}\BibitemShut {NoStop}%
\bibitem [{\citenamefont {Knaak}\ \emph {et~al.}(1986)\citenamefont {Knaak},
  \citenamefont {Hau{\ss}}, \citenamefont {Kummrow},\ and\ \citenamefont
  {Mei{\ss}ner}}]{knaak}%
  \BibitemOpen
  \bibfield  {author} {\bibinfo {author} {\bibfnamefont {W.}~\bibnamefont
  {Knaak}}, \bibinfo {author} {\bibfnamefont {T.}~\bibnamefont {Hau{\ss}}},
  \bibinfo {author} {\bibfnamefont {M.}~\bibnamefont {Kummrow}},\ and\ \bibinfo
  {author} {\bibfnamefont {M.}~\bibnamefont {Mei{\ss}ner}},\ }in\ \href
  {https://doi.org/10.1007/978-3-642-82912-3_52} {\emph {\bibinfo {booktitle}
  {Phonon Scattering in Condensed Matter V}}},\ \bibinfo {editor} {edited by\
  \bibinfo {editor} {\bibfnamefont {A.~C.}\ \bibnamefont {Anderson}}\ and\
  \bibinfo {editor} {\bibfnamefont {J.~P.}\ \bibnamefont {Wolfe}}}\ (\bibinfo
  {publisher} {Springer Berlin Heidelberg},\ \bibinfo {address} {Berlin,
  Heidelberg},\ \bibinfo {year} {1986})\ pp.\ \bibinfo {pages}
  {174--176}\BibitemShut {NoStop}%
\bibitem [{\citenamefont {H\"olzer}\ \emph {et~al.}(1997)\citenamefont
  {H\"olzer}, \citenamefont {Fritsch}, \citenamefont {Deutsch}, \citenamefont
  {H\"artwig},\ and\ \citenamefont {F\"orster}}]{fe55}%
  \BibitemOpen
  \bibfield  {author} {\bibinfo {author} {\bibfnamefont {G.}~\bibnamefont
  {H\"olzer}}, \bibinfo {author} {\bibfnamefont {M.}~\bibnamefont {Fritsch}},
  \bibinfo {author} {\bibfnamefont {M.}~\bibnamefont {Deutsch}}, \bibinfo
  {author} {\bibfnamefont {J.}~\bibnamefont {H\"artwig}},\ and\ \bibinfo
  {author} {\bibfnamefont {E.}~\bibnamefont {F\"orster}},\ }\bibinfo {title}
  {$K{\ensuremath{\alpha}}_{1,2}$ and $K{\ensuremath{\beta}}_{1,3}$ x-ray
  emission lines of the $3d$ transition metals},\ \href
  {https://doi.org/10.1103/PhysRevA.56.4554} {\bibfield  {journal} {\bibinfo
  {journal} {Phys. Rev. A}\ }\textbf {\bibinfo {volume} {56}},\ \bibinfo
  {pages} {4554} (\bibinfo {year} {1997})}\BibitemShut {NoStop}%
\bibitem [{\citenamefont {Schweppe}\ \emph {et~al.}(1994)\citenamefont
  {Schweppe}, \citenamefont {Deslattes}, \citenamefont {Mooney},\ and\
  \citenamefont {Powell}}]{alfluor}%
  \BibitemOpen
  \bibfield  {author} {\bibinfo {author} {\bibfnamefont {J.}~\bibnamefont
  {Schweppe}}, \bibinfo {author} {\bibfnamefont {R.~D.}\ \bibnamefont
  {Deslattes}}, \bibinfo {author} {\bibfnamefont {T.}~\bibnamefont {Mooney}},\
  and\ \bibinfo {author} {\bibfnamefont {C.~J.}\ \bibnamefont {Powell}},\
  }\bibinfo {title} {Accurate measurement of Mg and Al
  K${\ensuremath{\alpha}}_{1,2}$ X-ray energy profiles},\ \href
  {https://doi.org/10.1016/0368-2048(93)02059-U} {\bibfield  {journal}
  {\bibinfo  {journal} {J. Electron Spectrosc.}\ }\textbf {\bibinfo {volume}
  {67}},\ \bibinfo {pages} {463 } (\bibinfo {year} {1994})}\BibitemShut
  {NoStop}%
\bibitem [{\citenamefont {Hansen}\ \emph {et~al.}(2010)\citenamefont {Hansen},
  \citenamefont {DeJongh}, \citenamefont {Hall}, \citenamefont {Hines},
  \citenamefont {Huber}, \citenamefont {Kiper}, \citenamefont {Mandic},
  \citenamefont {Rau}, \citenamefont {Saab}, \citenamefont {Seitz} \emph
  {et~al.}}]{revc}%
  \BibitemOpen
  \bibfield  {author} {\bibinfo {author} {\bibfnamefont {S.}~\bibnamefont
  {Hansen}}, \bibinfo {author} {\bibfnamefont {F.}~\bibnamefont {DeJongh}},
  \bibinfo {author} {\bibfnamefont {J.}~\bibnamefont {Hall}}, \bibinfo {author}
  {\bibfnamefont {B.~A.}\ \bibnamefont {Hines}}, \bibinfo {author}
  {\bibfnamefont {M.~E.}\ \bibnamefont {Huber}}, \bibinfo {author}
  {\bibfnamefont {T.}~\bibnamefont {Kiper}}, \bibinfo {author} {\bibfnamefont
  {V.}~\bibnamefont {Mandic}}, \bibinfo {author} {\bibfnamefont
  {W.}~\bibnamefont {Rau}}, \bibinfo {author} {\bibfnamefont {T.}~\bibnamefont
  {Saab}}, \bibinfo {author} {\bibfnamefont {D.}~\bibnamefont {Seitz}}, \emph
  {et~al.},\ }in\ \href {https://doi.org/10.1109/NSSMIC.2010.5874000} {\emph
  {\bibinfo {booktitle} {IEEE Nuclear Science Symposium Medical Imaging
  Conference}}}\ (\bibinfo {address} {Knoxville, TN},\ \bibinfo {year} {2010})\
  pp.\ \bibinfo {pages} {1392--1395}\BibitemShut {NoStop}%
\bibitem [{\citenamefont {{Zadeh}}\ and\ \citenamefont
  {{Ragazzini}}(1952)}]{OF}%
  \BibitemOpen
  \bibfield  {author} {\bibinfo {author} {\bibfnamefont {L.~A.}\ \bibnamefont
  {{Zadeh}}}\ and\ \bibinfo {author} {\bibfnamefont {J.~R.}\ \bibnamefont
  {{Ragazzini}}},\ }\bibinfo {title} {Optimum Filters for the Detection of
  Signals in Noise},\ \href {https://doi.org/10.1109/JRPROC.1952.274117}
  {\bibfield  {journal} {\bibinfo  {journal} {Proc. IRE}\ }\textbf {\bibinfo
  {volume} {40}},\ \bibinfo {pages} {1223} (\bibinfo {year}
  {1952})}\BibitemShut {NoStop}%
\bibitem [{\citenamefont {Golwala}(2000)}]{golwala}%
  \BibitemOpen
  \bibfield  {author} {\bibinfo {author} {\bibfnamefont {S.~R.}\ \bibnamefont
  {Golwala}},\ }\emph {\bibinfo {title} {Exclusion limits on the WIMP nucleon
  elastic scattering cross-section from the Cryogenic Dark Matter Search}},\
  \href {https://doi.org/10.2172/1421437} {Ph.D. thesis},\ \bibinfo  {school}
  {University of California, Berkeley} (\bibinfo {year} {2000})\BibitemShut
  {NoStop}%
\bibitem [{\citenamefont {Reed}\ and\ \citenamefont {Ware}(1972)}]{Reed_1972}%
  \BibitemOpen
  \bibfield  {author} {\bibinfo {author} {\bibfnamefont {S.~J.~B.}\
  \bibnamefont {Reed}}\ and\ \bibinfo {author} {\bibfnamefont {N.~G.}\
  \bibnamefont {Ware}},\ }\bibinfo {title} {Escape peaks and internal
  fluorescence in X-ray spectra recorded with lithium drifted silicon
  detectors},\ \href {https://doi.org/10.1088/0022-3735/5/6/029} {\bibfield
  {journal} {\bibinfo  {journal} {J. Phys. E Sci. Instrum.}\ }\textbf {\bibinfo
  {volume} {5}},\ \bibinfo {pages} {582} (\bibinfo {year} {1972})}\BibitemShut
  {NoStop}%
\bibitem [{\citenamefont {De~Pascale}\ \emph {et~al.}(1993)\citenamefont
  {De~Pascale}, \citenamefont {Morselli}, \citenamefont {Picozza},
  \citenamefont {Golden}, \citenamefont {Grimani}, \citenamefont {Kimbell},
  \citenamefont {Stephens}, \citenamefont {Stochaj}, \citenamefont {Webber},
  \citenamefont {Basini} \emph {et~al.}}]{muonflux}%
  \BibitemOpen
  \bibfield  {author} {\bibinfo {author} {\bibfnamefont {M.~P.}\ \bibnamefont
  {De~Pascale}}, \bibinfo {author} {\bibfnamefont {A.}~\bibnamefont
  {Morselli}}, \bibinfo {author} {\bibfnamefont {P.}~\bibnamefont {Picozza}},
  \bibinfo {author} {\bibfnamefont {R.~L.}\ \bibnamefont {Golden}}, \bibinfo
  {author} {\bibfnamefont {C.}~\bibnamefont {Grimani}}, \bibinfo {author}
  {\bibfnamefont {B.~L.}\ \bibnamefont {Kimbell}}, \bibinfo {author}
  {\bibfnamefont {S.~A.}\ \bibnamefont {Stephens}}, \bibinfo {author}
  {\bibfnamefont {S.~J.}\ \bibnamefont {Stochaj}}, \bibinfo {author}
  {\bibfnamefont {W.~R.}\ \bibnamefont {Webber}}, \bibinfo {author}
  {\bibfnamefont {G.}~\bibnamefont {Basini}}, \emph {et~al.},\ }\bibinfo
  {title} {Absolute spectrum and charge ratio of cosmic ray muons in the energy
  region from 0.2 GeV to 100 GeV at 600 m above sea level},\ \href
  {https://doi.org/10.1029/92JA02672} {\bibfield  {journal} {\bibinfo
  {journal} {J. Geophys. Res.}\ }\textbf {\bibinfo {volume} {98}},\ \bibinfo
  {pages} {3501} (\bibinfo {year} {1993})}\BibitemShut {NoStop}%
\bibitem [{\citenamefont {Lewin}\ and\ \citenamefont
  {Smith}(1996)}]{lewinsmith}%
  \BibitemOpen
  \bibfield  {author} {\bibinfo {author} {\bibfnamefont {J.~D.}\ \bibnamefont
  {Lewin}}\ and\ \bibinfo {author} {\bibfnamefont {P.~F.}\ \bibnamefont
  {Smith}},\ }\bibinfo {title} {Review of mathematics, numerical factors, and
  corrections for dark matter experiments based on elastic nuclear recoil},\
  \href {https://doi.org/10.1016/S0927-6505(96)00047-3} {\bibfield  {journal}
  {\bibinfo  {journal} {Astropart. Phys.}\ }\textbf {\bibinfo {volume} {6}},\
  \bibinfo {pages} {87 } (\bibinfo {year} {1996})}\BibitemShut {NoStop}%
\bibitem [{\citenamefont {Kerr}\ and\ \citenamefont
  {Lynden-Bell}(1986)}]{kerr}%
  \BibitemOpen
  \bibfield  {author} {\bibinfo {author} {\bibfnamefont {F.~J.}\ \bibnamefont
  {Kerr}}\ and\ \bibinfo {author} {\bibfnamefont {D.}~\bibnamefont
  {Lynden-Bell}},\ }\bibinfo {title} {{Review of galactic constants}},\ \href
  {https://doi.org/10.1093/mnras/221.4.1023} {\bibfield  {journal} {\bibinfo
  {journal} {Mon. Not. R. Astron. Soc.}\ }\textbf {\bibinfo {volume} {221}},\
  \bibinfo {pages} {1023} (\bibinfo {year} {1986})}\BibitemShut {NoStop}%
\bibitem [{\citenamefont {Smith}\ \emph {et~al.}(2007)\citenamefont {Smith},
  \citenamefont {Ruchti}, \citenamefont {Helmi}, \citenamefont {Wyse},
  \citenamefont {Fulbright}, \citenamefont {Freeman}, \citenamefont {Navarro},
  \citenamefont {Seabroke}, \citenamefont {Steinmetz}, \citenamefont {Williams}
  \emph {et~al.}}]{rave}%
  \BibitemOpen
  \bibfield  {author} {\bibinfo {author} {\bibfnamefont {M.~C.}\ \bibnamefont
  {Smith}}, \bibinfo {author} {\bibfnamefont {G.~R.}\ \bibnamefont {Ruchti}},
  \bibinfo {author} {\bibfnamefont {A.}~\bibnamefont {Helmi}}, \bibinfo
  {author} {\bibfnamefont {R.~F.~G.}\ \bibnamefont {Wyse}}, \bibinfo {author}
  {\bibfnamefont {J.~P.}\ \bibnamefont {Fulbright}}, \bibinfo {author}
  {\bibfnamefont {K.~C.}\ \bibnamefont {Freeman}}, \bibinfo {author}
  {\bibfnamefont {J.~F.}\ \bibnamefont {Navarro}}, \bibinfo {author}
  {\bibfnamefont {G.~M.}\ \bibnamefont {Seabroke}}, \bibinfo {author}
  {\bibfnamefont {M.}~\bibnamefont {Steinmetz}}, \bibinfo {author}
  {\bibfnamefont {M.}~\bibnamefont {Williams}}, \emph {et~al.},\ }\bibinfo
  {title} {{The RAVE survey: constraining the local Galactic escape speed}},\
  \href {https://doi.org/10.1111/j.1365-2966.2007.11964.x} {\bibfield
  {journal} {\bibinfo  {journal} {Mon. Not. R. Astron. Soc.}\ }\textbf
  {\bibinfo {volume} {379}},\ \bibinfo {pages} {755} (\bibinfo {year}
  {2007})}\BibitemShut {NoStop}%
\bibitem [{\citenamefont {Schönrich}\ \emph {et~al.}(2010)\citenamefont
  {Schönrich}, \citenamefont {Binney},\ and\ \citenamefont
  {Dehnen}}]{schonrich}%
  \BibitemOpen
  \bibfield  {author} {\bibinfo {author} {\bibfnamefont {R.}~\bibnamefont
  {Schönrich}}, \bibinfo {author} {\bibfnamefont {J.}~\bibnamefont {Binney}},\
  and\ \bibinfo {author} {\bibfnamefont {W.}~\bibnamefont {Dehnen}},\ }\bibinfo
  {title} {{Local kinematics and the local standard of rest}},\ \href
  {https://doi.org/10.1111/j.1365-2966.2010.16253.x} {\bibfield  {journal}
  {\bibinfo  {journal} {Mon. Not. R. Astron. Soc.}\ }\textbf {\bibinfo {volume}
  {403}},\ \bibinfo {pages} {1829} (\bibinfo {year} {2010})}\BibitemShut
  {NoStop}%
\bibitem [{\citenamefont {Benoit}\ \emph {et~al.}(2007)\citenamefont {Benoit}
  \emph {et~al.}}]{BENOIT2007558}%
  \BibitemOpen
  \bibfield  {author} {\bibinfo {author} {\bibfnamefont {A.}~\bibnamefont
  {Benoit}} \emph {et~al.} (\bibinfo {collaboration} {EDELWEISS
  Collaboration}),\ }\bibinfo {title} {Measurement of the response of
  heat-and-ionization germanium detectors to nuclear recoils},\ \href
  {https://doi.org/10.1016/j.nima.2007.04.118} {\bibfield  {journal} {\bibinfo
  {journal} {Nucl. Instrum. Methods Phys. Res. A}\ }\textbf {\bibinfo {volume}
  {577}},\ \bibinfo {pages} {558 } (\bibinfo {year} {2007})}\BibitemShut
  {NoStop}%
\bibitem [{\citenamefont {Agnese}\ \emph
  {et~al.}(2018{\natexlab{b}})\citenamefont {Agnese} \emph {et~al.}}]{206Pb}%
  \BibitemOpen
  \bibfield  {author} {\bibinfo {author} {\bibfnamefont {R.}~\bibnamefont
  {Agnese}} \emph {et~al.} (\bibinfo {collaboration} {SuperCDMS
  Collaboration}),\ }\bibinfo {title} {Energy loss due to defect formation from
  $^{206}$Pb recoils in SuperCDMS germanium detectors},\ \href
  {https://doi.org/10.1063/1.5041457} {\bibfield  {journal} {\bibinfo
  {journal} {Appl. Phys. Lett.}\ }\textbf {\bibinfo {volume} {113}},\ \bibinfo
  {pages} {092101} (\bibinfo {year} {2018}{\natexlab{b}})}\BibitemShut
  {NoStop}%
\bibitem [{\citenamefont {Yellin}(2002)}]{yellin_upper}%
  \BibitemOpen
  \bibfield  {author} {\bibinfo {author} {\bibfnamefont {S.}~\bibnamefont
  {Yellin}},\ }\bibinfo {title} {Finding an upper limit in the presence of an
  unknown background},\ \href {https://doi.org/10.1103/PhysRevD.66.032005}
  {\bibfield  {journal} {\bibinfo  {journal} {Phys. Rev. D}\ }\textbf {\bibinfo
  {volume} {66}},\ \bibinfo {pages} {032005} (\bibinfo {year}
  {2002})}\BibitemShut {NoStop}%
\bibitem [{\citenamefont {Yellin}()}]{yellin_extended}%
  \BibitemOpen
  \bibfield  {author} {\bibinfo {author} {\bibfnamefont {S.}~\bibnamefont
  {Yellin}},\ }\href@noop {} {\bibinfo {title} {Extending the optimum interval
  method}},\ \Eprint {https://arxiv.org/abs/0709.2701} {arXiv:0709.2701}
  \BibitemShut {NoStop}%
\bibitem [{\citenamefont {Kavanagh}(2017)}]{verne}%
  \BibitemOpen
  \bibfield  {author} {\bibinfo {author} {\bibfnamefont {B.~J.}\ \bibnamefont
  {Kavanagh}},\ }\href@noop {} {\bibinfo {title} {bradkav/verne: Release}},\
  \bibinfo {howpublished} {\url{https://doi.org/10.5281/zenodo.1116305}}
  (\bibinfo {year} {2017})\BibitemShut {NoStop}%
\bibitem [{\citenamefont {Starkman}\ \emph {et~al.}(1990)\citenamefont
  {Starkman}, \citenamefont {Gould}, \citenamefont {Esmailzadeh},\ and\
  \citenamefont {Dimopoulos}}]{starkman1990}%
  \BibitemOpen
  \bibfield  {author} {\bibinfo {author} {\bibfnamefont {G.~D.}\ \bibnamefont
  {Starkman}}, \bibinfo {author} {\bibfnamefont {A.}~\bibnamefont {Gould}},
  \bibinfo {author} {\bibfnamefont {R.}~\bibnamefont {Esmailzadeh}},\ and\
  \bibinfo {author} {\bibfnamefont {S.}~\bibnamefont {Dimopoulos}},\ }\bibinfo
  {title} {Opening the window on strongly interacting dark matter},\ \href
  {https://doi.org/10.1103/PhysRevD.41.3594} {\bibfield  {journal} {\bibinfo
  {journal} {Phys. Rev. D}\ }\textbf {\bibinfo {volume} {41}},\ \bibinfo
  {pages} {3594} (\bibinfo {year} {1990})}\BibitemShut {NoStop}%
\bibitem [{\citenamefont {Zaharijas}\ and\ \citenamefont
  {Farrar}(2005)}]{zaharijas2005}%
  \BibitemOpen
  \bibfield  {author} {\bibinfo {author} {\bibfnamefont {G.}~\bibnamefont
  {Zaharijas}}\ and\ \bibinfo {author} {\bibfnamefont {G.~R.}\ \bibnamefont
  {Farrar}},\ }\bibinfo {title} {Window in the dark matter exclusion limits},\
  \href {https://doi.org/10.1103/PhysRevD.72.083502} {\bibfield  {journal}
  {\bibinfo  {journal} {Phys. Rev. D}\ }\textbf {\bibinfo {volume} {72}},\
  \bibinfo {pages} {083502} (\bibinfo {year} {2005})}\BibitemShut {NoStop}%
\bibitem [{\citenamefont {Emken}\ and\ \citenamefont
  {Kouvaris}(2018)}]{overburden}%
  \BibitemOpen
  \bibfield  {author} {\bibinfo {author} {\bibfnamefont {T.}~\bibnamefont
  {Emken}}\ and\ \bibinfo {author} {\bibfnamefont {C.}~\bibnamefont
  {Kouvaris}},\ }\bibinfo {title} {How blind are underground and surface
  detectors to strongly interacting dark matter?},\ \href
  {https://doi.org/10.1103/PhysRevD.97.115047} {\bibfield  {journal} {\bibinfo
  {journal} {Phys. Rev. D}\ }\textbf {\bibinfo {volume} {97}},\ \bibinfo
  {pages} {115047} (\bibinfo {year} {2018})}\BibitemShut {NoStop}%
\bibitem [{\citenamefont {Kavanagh}(2018)}]{PhysRevD.97.123013}%
  \BibitemOpen
  \bibfield  {author} {\bibinfo {author} {\bibfnamefont {B.~J.}\ \bibnamefont
  {Kavanagh}},\ }\bibinfo {title} {Earth scattering of superheavy dark matter:
  Updated constraints from detectors old and new},\ \href
  {https://doi.org/10.1103/PhysRevD.97.123013} {\bibfield  {journal} {\bibinfo
  {journal} {Phys. Rev. D}\ }\textbf {\bibinfo {volume} {97}},\ \bibinfo
  {pages} {123013} (\bibinfo {year} {2018})}\BibitemShut {NoStop}%
\bibitem [{\citenamefont {Aprile}\ \emph
  {et~al.}(2019{\natexlab{b}})\citenamefont {Aprile} \emph
  {et~al.}}]{PhysRevLett.123.241803}%
  \BibitemOpen
  \bibfield  {author} {\bibinfo {author} {\bibfnamefont {E.}~\bibnamefont
  {Aprile}} \emph {et~al.} (\bibinfo {collaboration} {XENON Collaboration}),\
  }\bibinfo {title} {Search for Light Dark Matter Interactions Enhanced by the
  Migdal Effect or Bremsstrahlung in XENON1T},\ \href
  {https://doi.org/10.1103/PhysRevLett.123.241803} {\bibfield  {journal}
  {\bibinfo  {journal} {Phys. Rev. Lett.}\ }\textbf {\bibinfo {volume} {123}},\
  \bibinfo {pages} {241803} (\bibinfo {year} {2019}{\natexlab{b}})}\BibitemShut
  {NoStop}%
\bibitem [{\citenamefont {Armengaud}\ \emph {et~al.}(2019)\citenamefont
  {Armengaud} \emph {et~al.}}]{edelweiss}%
  \BibitemOpen
  \bibfield  {author} {\bibinfo {author} {\bibfnamefont {E.}~\bibnamefont
  {Armengaud}} \emph {et~al.} (\bibinfo {collaboration} {EDELWEISS
  Collaboration}),\ }\bibinfo {title} {Searching for low-mass dark matter
  particles with a massive Ge bolometer operated above ground},\ \href
  {https://doi.org/10.1103/PhysRevD.99.082003} {\bibfield  {journal} {\bibinfo
  {journal} {Phys. Rev. D}\ }\textbf {\bibinfo {volume} {99}},\ \bibinfo
  {pages} {082003} (\bibinfo {year} {2019})}\BibitemShut {NoStop}%
\bibitem [{\citenamefont {Abdelhameed}\ \emph {et~al.}(2019)\citenamefont
  {Abdelhameed} \emph {et~al.}}]{cresst2019}%
  \BibitemOpen
  \bibfield  {author} {\bibinfo {author} {\bibfnamefont {A.~H.}\ \bibnamefont
  {Abdelhameed}} \emph {et~al.} (\bibinfo {collaboration} {CRESST
  Collaboration}),\ }\bibinfo {title} {First results from the CRESST-III
  low-mass dark matter program},\ \href
  {https://doi.org/10.1103/PhysRevD.100.102002} {\bibfield  {journal} {\bibinfo
   {journal} {Phys. Rev. D}\ }\textbf {\bibinfo {volume} {100}},\ \bibinfo
  {pages} {102002} (\bibinfo {year} {2019})}\BibitemShut {NoStop}%
\bibitem [{\citenamefont {Angloher}\ \emph {et~al.}(2017)\citenamefont
  {Angloher} \emph {et~al.}}]{cresst_aboveground}%
  \BibitemOpen
  \bibfield  {author} {\bibinfo {author} {\bibfnamefont {G.}~\bibnamefont
  {Angloher}} \emph {et~al.} (\bibinfo {collaboration} {CRESST
  Collaboration}),\ }\bibinfo {title} {Results on MeV-scale dark matter from a
  gram-scale cryogenic calorimeter operated above ground},\ \href
  {https://doi.org/10.1140/epjc/s10052-017-5223-9} {\bibfield  {journal}
  {\bibinfo  {journal} {Eur. Phys. J. C}\ }\textbf {\bibinfo {volume} {77}},\
  \bibinfo {pages} {637} (\bibinfo {year} {2017})}\BibitemShut {NoStop}%
\bibitem [{\citenamefont {Aguilar-Arevalo}\ \emph {et~al.}(2020)\citenamefont
  {Aguilar-Arevalo} \emph {et~al.}}]{PhysRevLett.125.241803}%
  \BibitemOpen
  \bibfield  {author} {\bibinfo {author} {\bibfnamefont {A.}~\bibnamefont
  {Aguilar-Arevalo}} \emph {et~al.} (\bibinfo {collaboration} {DAMIC
  Collaboration}),\ }\bibinfo {title} {Results on Low-Mass Weakly Interacting
  Massive Particles from an 11 kg d Target Exposure of DAMIC at SNOLAB},\ \href
  {https://doi.org/10.1103/PhysRevLett.125.241803} {\bibfield  {journal}
  {\bibinfo  {journal} {Phys. Rev. Lett.}\ }\textbf {\bibinfo {volume} {125}},\
  \bibinfo {pages} {241803} (\bibinfo {year} {2020})}\BibitemShut {NoStop}%
\bibitem [{\citenamefont {Angloher}\ \emph {et~al.}(2016)\citenamefont
  {Angloher} \emph {et~al.}}]{cresst2}%
  \BibitemOpen
  \bibfield  {author} {\bibinfo {author} {\bibfnamefont {G.}~\bibnamefont
  {Angloher}} \emph {et~al.} (\bibinfo {collaboration} {CRESST
  Collaboration}),\ }\bibinfo {title} {Results on light dark matter particles
  with a low-threshold CRESST-II detector},\ \href
  {https://doi.org/10.1140/epjc/s10052-016-3877-3} {\bibfield  {journal}
  {\bibinfo  {journal} {Eur. Phys. J. C}\ }\textbf {\bibinfo {volume} {76}},\
  \bibinfo {pages} {25} (\bibinfo {year} {2016})}\BibitemShut {NoStop}%
\bibitem [{\citenamefont {Arnaud}\ \emph {et~al.}(2018)\citenamefont {Arnaud}
  \emph {et~al.}}]{newsg}%
  \BibitemOpen
  \bibfield  {author} {\bibinfo {author} {\bibfnamefont {Q.}~\bibnamefont
  {Arnaud}} \emph {et~al.} (\bibinfo {collaboration} {NEWS-G Collaboration}),\
  }\bibinfo {title} {First results from the NEWS-G direct dark matter search
  experiment at the LSM},\ \href
  {https://doi.org/10.1016/j.astropartphys.2017.10.009} {\bibfield  {journal}
  {\bibinfo  {journal} {Astropart. Phys.}\ }\textbf {\bibinfo {volume} {97}},\
  \bibinfo {pages} {54 } (\bibinfo {year} {2018})}\BibitemShut {NoStop}%
\bibitem [{\citenamefont {Collar}(2018)}]{PhysRevD.98.023005}%
  \BibitemOpen
  \bibfield  {author} {\bibinfo {author} {\bibfnamefont {J.~I.}\ \bibnamefont
  {Collar}},\ }\bibinfo {title} {Search for a nonrelativistic component in the
  spectrum of cosmic rays at Earth},\ \href
  {https://doi.org/10.1103/PhysRevD.98.023005} {\bibfield  {journal} {\bibinfo
  {journal} {Phys. Rev. D}\ }\textbf {\bibinfo {volume} {98}},\ \bibinfo
  {pages} {023005} (\bibinfo {year} {2018})}\BibitemShut {NoStop}%
\bibitem [{\citenamefont {Digman}\ \emph {et~al.}(2019)\citenamefont {Digman},
  \citenamefont {Cappiello}, \citenamefont {Beacom}, \citenamefont {Hirata},\
  and\ \citenamefont {Peter}}]{PhysRevD.100.063013}%
  \BibitemOpen
  \bibfield  {author} {\bibinfo {author} {\bibfnamefont {M.~C.}\ \bibnamefont
  {Digman}}, \bibinfo {author} {\bibfnamefont {C.~V.}\ \bibnamefont
  {Cappiello}}, \bibinfo {author} {\bibfnamefont {J.~F.}\ \bibnamefont
  {Beacom}}, \bibinfo {author} {\bibfnamefont {C.~M.}\ \bibnamefont {Hirata}},\
  and\ \bibinfo {author} {\bibfnamefont {A.~H.~G.}\ \bibnamefont {Peter}},\
  }\bibinfo {title} {Not as big as a barn: Upper bounds on dark matter-nucleus
  cross sections},\ \href {https://doi.org/10.1103/PhysRevD.100.063013}
  {\bibfield  {journal} {\bibinfo  {journal} {Phys. Rev. D}\ }\textbf {\bibinfo
  {volume} {100}},\ \bibinfo {pages} {063013} (\bibinfo {year}
  {2019})}\BibitemShut {NoStop}%
\bibitem [{\citenamefont {\r{A}str\"om}\ \emph {et~al.}(2006)\citenamefont
  {\r{A}str\"om}, \citenamefont {{Di Stefano}}, \citenamefont {Pr\"obst},
  \citenamefont {Stodolsky}, \citenamefont {Timonen}, \citenamefont {Bucci},
  \citenamefont {Cooper}, \citenamefont {Cozzini}, \citenamefont {Feilitzsch},
  \citenamefont {Kraus} \emph {et~al.}}]{ASTROM2006262}%
  \BibitemOpen
  \bibfield  {author} {\bibinfo {author} {\bibfnamefont {J.}~\bibnamefont
  {\r{A}str\"om}}, \bibinfo {author} {\bibfnamefont {P.~C.~F.}\ \bibnamefont
  {{Di Stefano}}}, \bibinfo {author} {\bibfnamefont {F.}~\bibnamefont
  {Pr\"obst}}, \bibinfo {author} {\bibfnamefont {L.}~\bibnamefont {Stodolsky}},
  \bibinfo {author} {\bibfnamefont {J.}~\bibnamefont {Timonen}}, \bibinfo
  {author} {\bibfnamefont {C.}~\bibnamefont {Bucci}}, \bibinfo {author}
  {\bibfnamefont {S.}~\bibnamefont {Cooper}}, \bibinfo {author} {\bibfnamefont
  {C.}~\bibnamefont {Cozzini}}, \bibinfo {author} {\bibfnamefont {F.~v.}\
  \bibnamefont {Feilitzsch}}, \bibinfo {author} {\bibfnamefont
  {H.}~\bibnamefont {Kraus}}, \emph {et~al.},\ }\bibinfo {title} {Fracture
  processes observed with a cryogenic detector},\ \href
  {https://doi.org/10.1016/j.physleta.2006.03.059} {\bibfield  {journal}
  {\bibinfo  {journal} {Phys. Lett. A}\ }\textbf {\bibinfo {volume} {356}},\
  \bibinfo {pages} {262 } (\bibinfo {year} {2006})}\BibitemShut {NoStop}%
\bibitem [{\citenamefont {Kurinsky}\ \emph {et~al.}(2020)\citenamefont
  {Kurinsky}, \citenamefont {Baxter}, \citenamefont {Kahn},\ and\ \citenamefont
  {Krnjaic}}]{PhysRevD.102.015017}%
  \BibitemOpen
  \bibfield  {author} {\bibinfo {author} {\bibfnamefont {N.}~\bibnamefont
  {Kurinsky}}, \bibinfo {author} {\bibfnamefont {D.}~\bibnamefont {Baxter}},
  \bibinfo {author} {\bibfnamefont {Y.}~\bibnamefont {Kahn}},\ and\ \bibinfo
  {author} {\bibfnamefont {G.}~\bibnamefont {Krnjaic}},\ }\bibinfo {title}
  {Dark matter interpretation of excesses in multiple direct detection
  experiments},\ \href {https://doi.org/10.1103/PhysRevD.102.015017} {\bibfield
   {journal} {\bibinfo  {journal} {Phys. Rev. D}\ }\textbf {\bibinfo {volume}
  {102}},\ \bibinfo {pages} {015017} (\bibinfo {year} {2020})}\BibitemShut
  {NoStop}%
\bibitem [{\citenamefont {Hochberg}\ \emph {et~al.}(2016)\citenamefont
  {Hochberg}, \citenamefont {Pyle}, \citenamefont {Zhao},\ and\ \citenamefont
  {Zurek}}]{Hochberg_2016}%
  \BibitemOpen
  \bibfield  {author} {\bibinfo {author} {\bibfnamefont {Y.}~\bibnamefont
  {Hochberg}}, \bibinfo {author} {\bibfnamefont {M.}~\bibnamefont {Pyle}},
  \bibinfo {author} {\bibfnamefont {Y.}~\bibnamefont {Zhao}},\ and\ \bibinfo
  {author} {\bibfnamefont {K.~M.}\ \bibnamefont {Zurek}},\ }\bibinfo {title}
  {Detecting superlight dark matter with Fermi-degenerate materials},\ \href
  {https://doi.org/10.1007/jhep08(2016)057} {\bibfield  {journal} {\bibinfo
  {journal} {J. High Energ. Phys.}\ }\textbf {\bibinfo {volume} {2016}},\
  \bibinfo {pages} {57} (\bibinfo {year} {2016})}\BibitemShut {NoStop}%
\bibitem [{\citenamefont {Knapen}\ \emph {et~al.}(2018)\citenamefont {Knapen},
  \citenamefont {Lin}, \citenamefont {Pyle},\ and\ \citenamefont
  {Zurek}}]{KNAPEN2018386}%
  \BibitemOpen
  \bibfield  {author} {\bibinfo {author} {\bibfnamefont {S.}~\bibnamefont
  {Knapen}}, \bibinfo {author} {\bibfnamefont {T.}~\bibnamefont {Lin}},
  \bibinfo {author} {\bibfnamefont {M.}~\bibnamefont {Pyle}},\ and\ \bibinfo
  {author} {\bibfnamefont {K.~M.}\ \bibnamefont {Zurek}},\ }\bibinfo {title}
  {Detection of light dark matter with optical phonons in polar materials},\
  \href {https://doi.org/10.1016/j.physletb.2018.08.064} {\bibfield  {journal}
  {\bibinfo  {journal} {Phys. Lett. B}\ }\textbf {\bibinfo {volume} {785}},\
  \bibinfo {pages} {386 } (\bibinfo {year} {2018})}\BibitemShut {NoStop}%
\end{thebibliography}
\end{document}